# Carbon Doping of $MgB_2$ by Toluene and Malic-Acid-in-Toluene

S. D. Bohnenstiehl[1], M. A. Susner[1], Y. Yang[1], E. W. Collings[1], M. D. Sumption[1], M. A. Rindfleisch[2], R. Boone[2]

[1]Center for Superconducting and Magnetic Materials, Department of Materials Science and Engineering, The Ohio State University, Columbus, OH, USA

[2]Hypertech Research, Columbus OH, USA

**Abstract**

The decomposition of malic acid in the presence of Mg and B was studied using Differential Scanning Calorimetry (DSC) and Thermogravimetric Analysis (TGA) which revealed that malic acid reacted with Mg but not B. Also, the addition of toluene to dissolve malic acid followed by subsequent drying resulted in no reaction with Mg, indicating that the malic acid had decomposed during the dissolution/drying stage. The total carbon contributed by toluene versus a toluene/5 wt% malic acid mixture was measured using a LECO CS600 carbon analyzer. The toluene sample contained ~0.4 wt% C while the toluene/malic acid mixture had ~1.5 wt% C, demonstrating that the toluene contributed a significant amount of carbon to the final product. Resistivity measurements on powder-in-tube $MgB_2$ monofilamentary wires established that the toluene/malic acid doped sample had the highest $B_{c2}$. However, the toluene-only sample had the highest transport $J_c$ over most of the magnetic field range (0-9 T), equaled only by that of toluene/malic acid sample in fields above 9 T.

**Introduction**

One of the challenges in $MgB_2$ research continues to be homogeneous doping of bulk and powder-in-tube (PIT) samples to obtain higher $B_{c2}$. Carbon is the only element known to substitute on the boron site in the $MgB_2$ structure. One method of obtaining carbon doped samples is to add a precursor to the magnesium/boron powder mixture that is reduced by magnesium at higher temperatures. Perhaps the most common $MgB_2$ additive is nanometer size SiC powder which is reduced by magnesium to form $Mg_2Si$ and carbon [1] which is then available for substitution on the boron site during $MgB_2$ formation.

Another successful carbon-containing additive is malic acid [2-6]. It is important to note that the malic acid is almost never added directly to the dry magnesium/boron powder mixture; instead, it is typically added to a toluene/boron powder slurry that is heated to ~150 °C under vacuum [2]. The dried mixture is then mixed with magnesium powder and processed into bulk or PIT samples. This method has been very effective at increasing transport properties via carbon doping; however, it is useful to determine how much the toluene itself may be contributing to the doping process since toluene alone has been shown to increase transport $J_c$ [7-8]. While toluene will not contribute more C than malic acid for the higher malic acid addition levels common in strand optimization studies [6], it is important to quantify this contribution. Such quantification is the focus of this study, rather than $J_c$ optimization per se. Given this objective, lower levels of malic acid were used to more easily discern the various contributions.

In this study, we used differential scanning calorimetry (DSC) and thermogravimetric analysis (TGA) to study thermal events accompanying the heating of (i) malic acid (ii) malic acid plus boron (50:50 by weight) and (iii) malic acid plus magnesium (50:50 by weight) to determine the events that occurred in the absence of toluene. Then, we performed DSC and TGA on a B/Mg/malic acid (5 wt %) mixture, using the toluene procedure as noted above, to study the change in thermal events caused by toluene. We made resistive measurements of the critical fields, $B_{irr}$ and $B_{c2}$, and transport measurements of critical current density, $J_c$, on PIT wires prepared from binary magnesium and boron powders and those doped with (i) toluene alone and (ii) toluene/malic acid (5 wt%) using the procedure in Kim et al. [2]. In addition we measured the total carbon content of each powder batch in order to compare the contributions due to toluene and malic acid, respectively. While these transport results, which were made on strands using a lower grade B, are not as high as we typically measure with amorphous B [9], the baseline contribution of toluene can be seen.

**Experimental Procedure**

All DSC and TGA samples were prepared from magnesium powder (99% pure), Tangshan boron powder (98% pure), and malic acid powder (99% pure). DSC measurements were collected with a TA Instruments 2920 under argon flow (99.998%). The temperature signal and heat flow were checked using the melting point and heat of fusion of indium for low temperature runs and magnesium for high temperature runs. Standard aluminum pans and lids were used for low temperature runs. High temperature runs above 600 °C used custom niobium pans of the kind used in a previous DSC investigation [10]. The niobium pans were similar in size and shape to the aluminum pans and lids. TGA data were collected with a Perkin-Elmer TGA 7 using a platinum pan. Furnace ramp rates for DSC and TGA were 10 °C/minute. The boron/malic acid and magnesium/malic acid mixtures used for DSC and TGA were dry mixed at approximately 50:50 wt%.

In order to test the effect of toluene on carbon doping levels, two separate batches of boron were made: (1) boron with malic acid and toluene, and (2) boron with toluene only. The malic acid doped boron was in the ratio of 1 gram malic acid to 20 grams boron. The processing procedure was to mix the malic acid and boron in 60 mL of toluene under argon in a glass vessel for one hour. After mixing, the vessel was transferred to a stainless steel retort which was heated to 155 °C for 3 hours under vacuum to remove all volatile organics. The toluene/boron batch followed the same procedure except no malic acid was included.

A two-gram sample of each of these batches was sent to LECO Corporation for carbon analysis along with a two-gram sample of untreated boron powder. A CS600 carbon analyzer was used for carbon analysis and was calibrated with 4 standards containing 1.00%, 0.53%, 0.13% and 0.023% carbon in weight percent. The untreated and toluene-only samples were measured twice and the toluene/malic-acid sample three times to establish repeatability. These measurements were made before adding magnesium powder, but fortunately carbon is not a common impurity for magnesium since there is no solubility for carbon in magnesium.

These three powders (boron, boron/toluene, boron/toluene/malic acid) were then mixed with Mg powder at a ratio appropriate for stoichiometric $MgB_2$ and made into PIT monofilamentary wires at HyperTech Research, Inc. In preparation for critical current density measurements, 10 cm lengths of each specimen were heat treated at 700 °C for 30 minutes under argon followed by a furnace cool. A 3 cm segment was cut from the center of each of the reacted strands for $I_c$ measurement. The voltage taps were 5 mm apart, and the voltage criterion was 1 µV/cm. Transport $J_c$ was measured at 4.2 K in transverse magnetic fields up to 10 T. For critical field determination, resistivity was measured as a function of temperature in transverse applied fields up to 14 T using a Quantum Design Model 6000 Physical Property Measurement System

(PPMS). The four point technique was applied to 1 cm samples using a sensing current of 5 mA and voltage tap separation of 5 mm. Contacts were attached using copper wire and silver paste from Ted Pella Inc. Criteria of 5% and 95% of the normal state resistivity were used to determine $B_{irr}$ and $B_{c2}$, respectively.

**Results and Discussion**

The DSC scan of malic acid by itself shows a typical melting endotherm at 130 °C followed by a wide decomposition endotherm from 185-275 °C (Fig. 1). The wide decomposition range may be due to a series of reactions that overlap at different temperatures. The TGA scan on malic acid powder also shows a wide decomposition range that ends at 275 °C, matching the DSC result. In general, these results support the use of 150 °C for vacuum drying of boron/toluene/malic acid mixtures since decomposition temperatures should be lowered in a vacuum environment assuming the toluene does not substantially change the kinetics of decomposition. Similar reactions take place after the malic acid is mixed with boron (Fig. 2). The weight loss percentage uses the amount of malic acid present in the starting mixture to calculate the percent weight loss. The boron is most likely inert with respect to malic acid given that there is no significant difference between the DSC and TGA data for malic acid by itself and boron mixed with malic acid.

The mixture of magnesium and malic acid has a completely different response (Fig. 3). A violent exothermic reaction occurs almost immediately after the onset of the malic acid melting endotherm at 130 °C. The TGA data also has a step change at this temperature which indicates the reaction generates some volatile products resulting in weight loss. Following this reaction, there is a broad decomposition endotherm and attendant weight loss. However, the total weight loss of malic acid is limited to approximately 50%.

A DSC scan of the dried boron/toluene/malic acid mixture that was mixed with Mg has none of the above thermal events at low temperatures (Fig. 4). This supports the assertion that the malic acid decomposed at ~150 °C under vacuum [3]. If any significant amount of malic acid was retained in the mixture after drying, then the Mg should have reacted with it at low temperature, as described above. The strong exothermic event at ~537 °C is typical of Mg and B powder mixtures and results in partial formation of $MgB_2$ [11-12]. The endothermic event at 650 °C is due to Mg melting which indicates an incomplete reaction at lower temperatures.

The LECO carbon analysis results establish that there is a carbon contribution from the toluene alone without any malic acid addition (Table 1). Although the carbon amount is not large (~0.4 wt %), these results suggest that toluene must be reacting or decomposing to leave a residue in some fashion. This is interesting in light of the fact that toluene and many other organics have been used in the past to increase transport $J_c$ in $MgB_2$ by increasing $B_{c2}$ through carbon doping [7-8]. The addition of 5 wt% malic acid increases the carbon content to ~1.5 wt% before mixing with Mg. If one assumes that these carbon amounts are additive, then it follows that ~1.1 wt% is from the malic acid and toluene makes up the balance (~0.4 wt %). Our batch composition of 1 gram malic acid to 20 grams boron gives a theoretical carbon addition of ~1.8 wt% assuming all the carbon in the malic acid is left behind. Thus, approximately 40% of the carbon in the malic acid is lost during decomposition, possibly in the form of CO and $CO_2$. We should note that the analyzed carbon levels reported here correspond to the total carbon in the sample and may not correspond to the actual amount of carbon that enters the boron sub-lattice during reaction. Also, these low levels of carbon are difficult to detect using lattice parameter measurements of the reacted $MgB_2$, which is why we used the LECO CS600 for our measurements.

The resistivity measurements indicate an increase in $B_{irr}$ and $B_{c2}$ at lower temperatures for both doped wires, with a larger increase seen for the toluene/malic acid wire (Fig. 5). This is as expected given that the toluene + malic acid wire contained ~1.5 wt% carbon instead of ~0.4 wt% for the toluene-only wire. However, the transport $J_c$ measurements show that the toluene-only wire had the highest transport $J_c$ over most of the field range (0-9 T), being approached by the toluene/malic acid doped strand at 10 T (Fig. 6). Of course, the malic acid doping level present in these wires is much lower than for optimized strands. Thus for the malic doping levels commonly used (~10 wt%), we would expect higher properties not only in the critical fields as shown in Figure 5, but also higher transport properties at high field as shown in Figure 6. However, it is interesting to compare the influence of the low level malic acid addition with the toluene addition.

We do see that both doped strands are approaching the undoped strand as the field is reduced, and based on the slopes, an extrapolation to zero field would be expected to lead to the highest $J_c$s for the undoped strand. This overall observation no doubt has a contribution due to the fact that $B_{c2}$ is increased for the doped strands [2]. However, we note that the toluene + malic acid strand has a reduced low field $J_c$ as compared to the toluene-only strand. The fact that the doped wires both approach the same transport $J_c$ at high field suggests that carbon doping was successful at some level for both strands. Focusing for a moment on the high field response, we note that if one assumes that all the carbon measured by the LECO analysis substituted for the boron, then the toluene/malic acid wire should have a substantially higher transport $J_c$ at high field than the toluene wire. Thus, either not all of the carbon from the malic acid is substituting on the boron sub-lattice, or alternatively, the transport is reduced by some other mechanism. The fact that the malic acid + toluene strand has a higher $B_{c2}$ value suggests that it is the latter. Similarly, the low field differences in $J_c$ may be due to either a greater $B_{c2}$ increase for the toluene + malic acid strand as seen in Fig. 5, or additional residual phases from the decomposition of the malic acid. Indeed, the addition of dopants is known to reduce connectivity in $MgB_2$ strands as compared to undoped strands [13].

In any case, it is of strong interest to note the practical aspects which are (i) the toluene adds its own carbon even though its intended function is only to act as a carrier for the malic acid, and (ii) the final influence on transport results is not a function of raw C-doping power alone, at least at these low levels of malic acid addition.

**Conclusions**

It has been shown that toluene, which is typically used as a carrier in the malic acid doping process, makes its own carbon contribution to the process which we have quantified. The present study used relatively low levels of malic acid doping to enable clear comparison to the toluene contribution. While the $B_{c2}$ and $B_{irr}$ enhancements responded more or less directly to the total amount of carbon in the wire, the strand with only toluene seemed to have a more efficient increase in transport $J_c$ per wt% C than the toluene + malic acid strand. The reason for this is not clear, but may be associated with secondary residual phases and connectivity issues. While further studies with optimized strand designs and higher levels of malic acid doping are needed, it is clearly important to know the total carbon content of the powder before reaction and the contributions from all potential sources of carbon in the doping process.

**Acknowledgements**

This work was supported by the U.S. Department of Energy, Office of High Energy Physics under Grant Number DE-FG02-95ER40900.

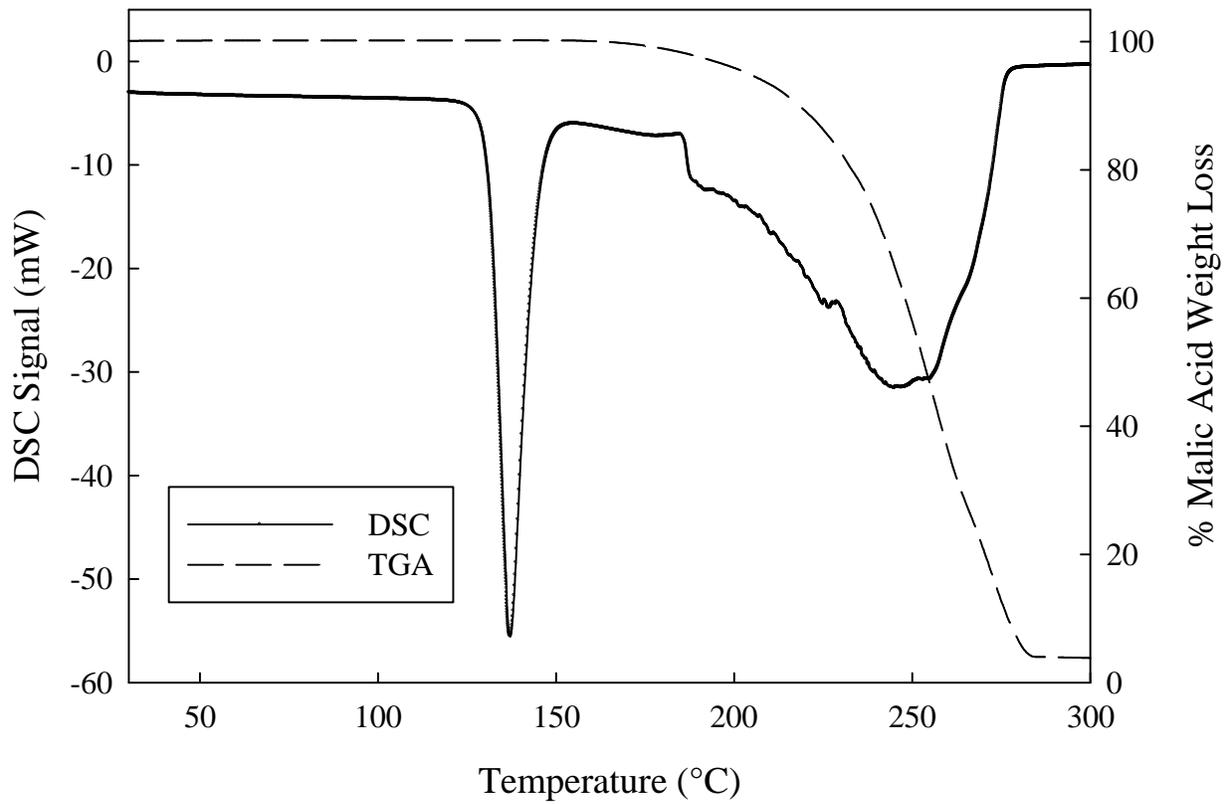

Figure 1. DSC and TGA on Malic Acid (Endothermic Down)

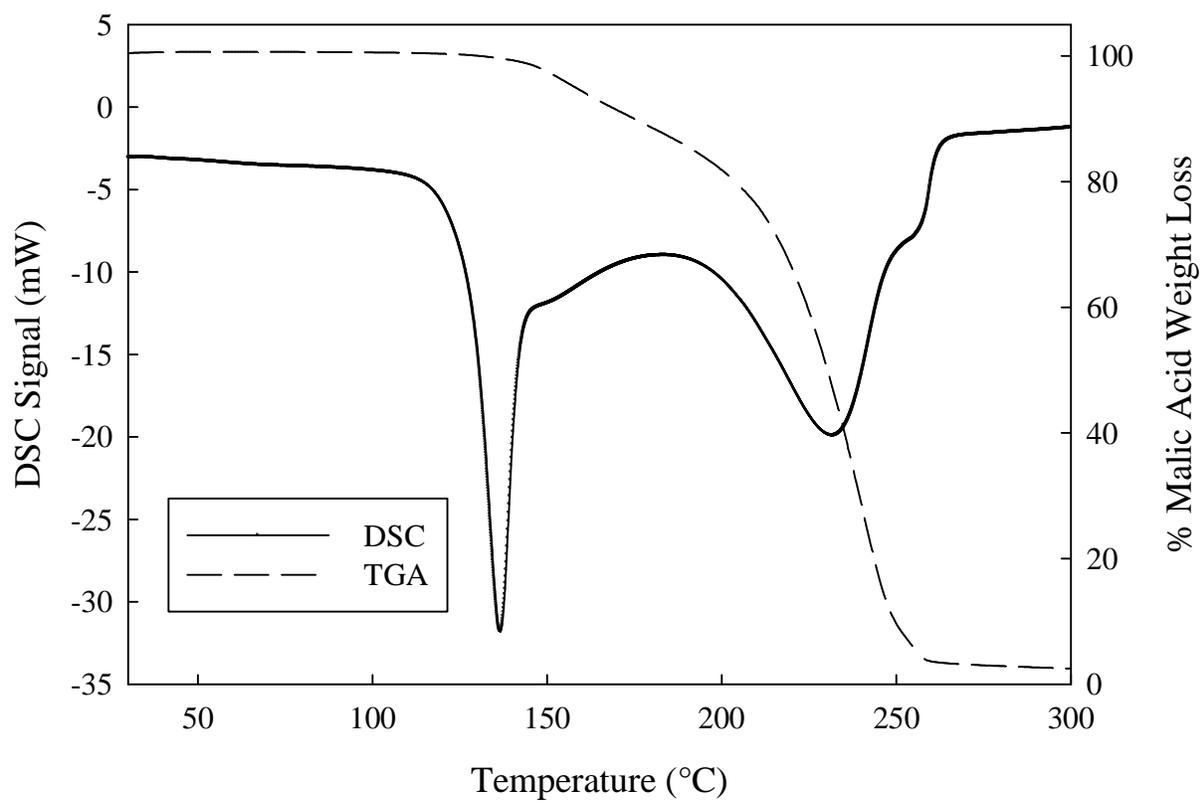

Figure 2. DSC and TGA on 50:50 wt% Boron/Malic Acid Mixture (Endothermic Down)

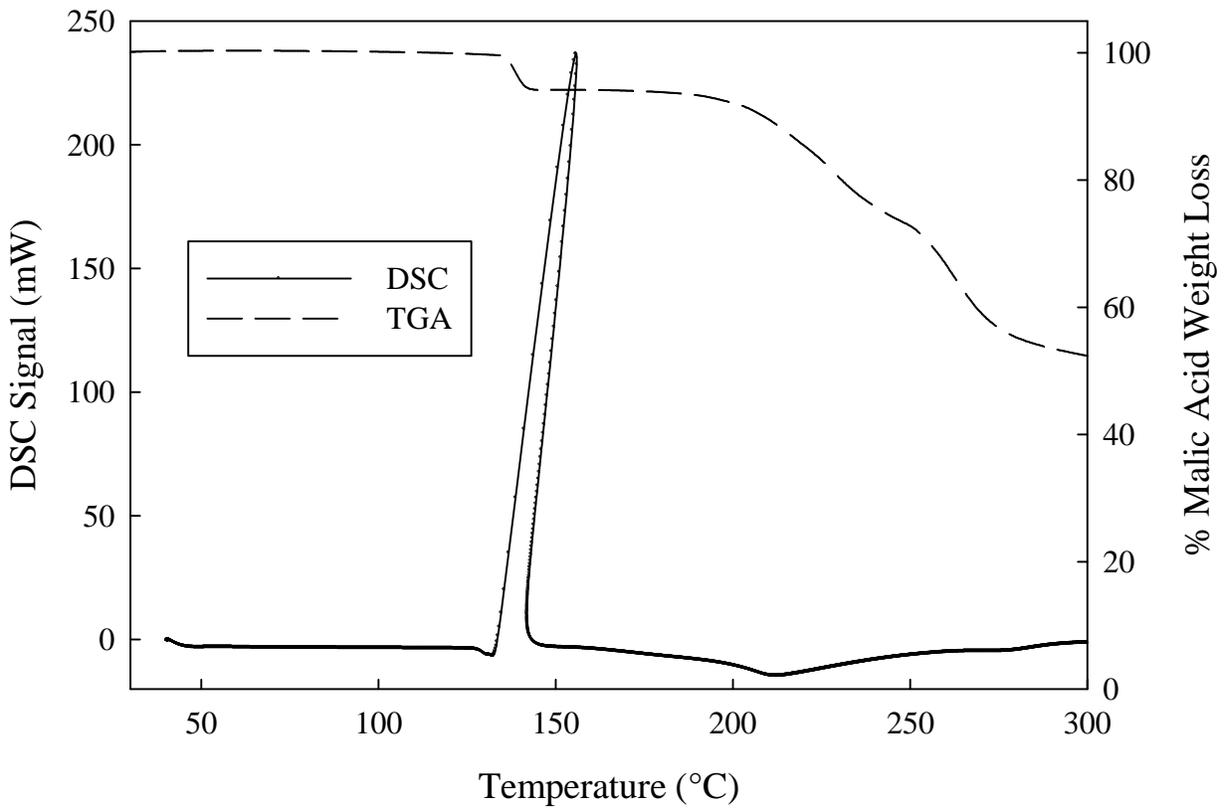

Figure 3. DSC and TGA on 50:50 wt% Magnesium/Malic Acid Mixture (Endothermic Down)

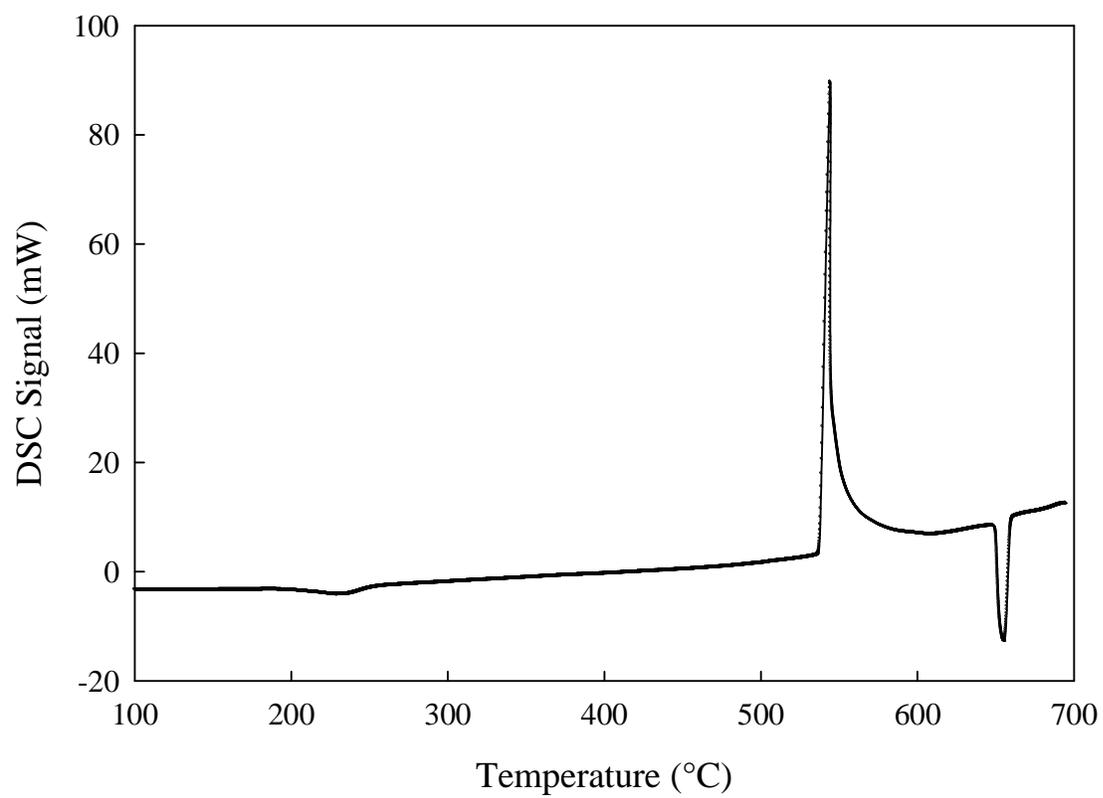

Figure 4. DSC on Mg + B mixture treated with toluene + malic acid (Endothermic Down)

| Boron – Untreated | Boron - Toluene | Boron - Toluene & Malic Acid |
|---|---|---|
| 0.077, 0.077 | 0.423, 0.437 | 1.47, 1.53, 1.44 |

Table 1. Carbon Concentration for Boron Powders in Weight Percent

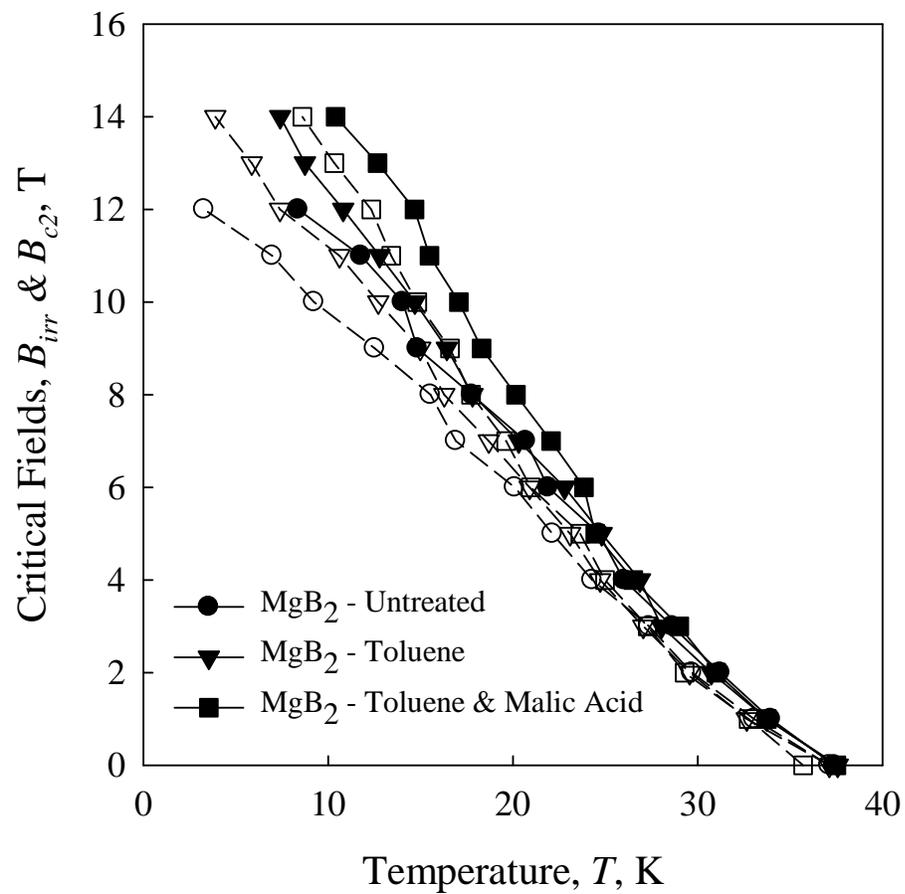

Figure 5. $B_{irr}$ (unfilled) and $B_{c2}$ (filled) of MgB$_2$ PIT Wires

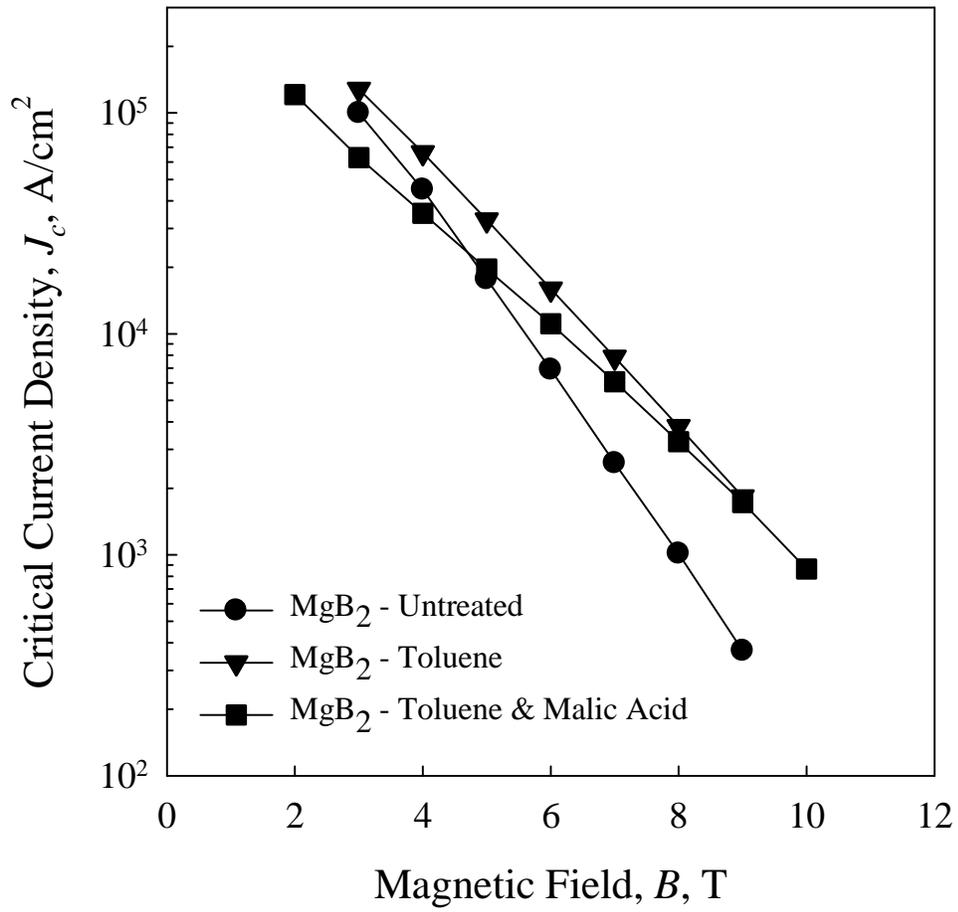

Figure 6. Transport $J_c$ on $MgB_2$ PIT Wires